\def\imo{i}
 \def\e{\epsilon}

\def\imo{i}
\def\re#1{Re(#1)}

\def\imo{i}
\def\rem#1{}

\def\be{\begin{equation}}
\def\ee{\end{equation}}
\def\bea{\begin{eqnarray}}
\def\eea{\end{eqnarray}}

\def\imo{i}

\let\pd\partial
\documentclass[11pt]{article}
\usepackage[dvips]{graphicx}
\usepackage{amsmath,amssymb,bm}
\usepackage{color}

\begin{document}
\title{Superluminal neutrinos and the tachyon`s stability in the rotating Universe}
\author{R. A. Konoplya\footnote{Email:konoplya\underline{~}roma@yahoo.com} \\
DAMTP, Centre for Mathematical Sciences, University of Cambridge, \\Wilberforce Road, Cambridge CB3 0WA, UK}

\date{}

\maketitle

\thispagestyle{empty}

\begin{abstract}
It is well-known that a hypothetical particle which moves faster than the light, a \emph{tachyon}, is unstable in
the Minkowski space-time. Here we shall show that, contrary to the Minkowski case, the tachyon is stable in the rotating Universe
described by a family of the G\"{o}del-like solutions, unless the absolute value of its mass is larger than some small constant which
is related to the universe`s rotation scale and is many orders less than the electron`s mass.
The smallness of this upper bound on the tachyon`s mass might be an explanation why we do not observe heavy tachyons.
Mathematically, the stability bound is similar to the well-known Breitenlohner-Freedman bound for the asymptotically anti-de Sitter (AdS) space-times.
\end{abstract}

\section{Introduction}

Tachyons are used in quantum field theory in order to describe the spontaneous symmetry breaking: the existence of the tachyons in the field`s spectrum implies the onset of instability of the vacuum state of the field \cite{Peskin}. In string theory, in a similar fashion, a
tachyon mode in the spectrum of the open sting means the instability of the D-brane to which the string is attached, so that the system
decays to a stable set of closed strings or D-branes (or both) \cite{Sen:1998sm}. Special attention is paid to tachyons in cosmology \cite{Narlikar}, \cite{Gorini:2003wa}, where the notion of the tachyonic matter is used when solving various problems connected to possible existence of dark matter.
In addition, one of the most intriguing issues is possible tachyonic structure of neutrinos, which has been discussed for a long time \cite{Chodosa}, and is generating interest nowadays \cite{Alfaro:2004aa}, in connection with the recent OPERA experiment \cite{OPERA:2011zb},
whose data indicates that the neutrinos move at a velocity, which is $2 \cdot 10^{-5} c $ above the speed of light $c$.
The experiment is attracting interest of theorists on a number of reasons \cite{Gubser:2011mp}-\cite{Pfeifer:2011ve}.
Nevertheless, the experiment announced recently by CERN is waiting for independent verification by other groups. The problem of tachyonic matter is interesting also by itself from the theoretical point of view, as it raises a number of challenging questions, such as problems with the space-time causality or quantization.
Yet, one of the insuperable problems is the above mentioned instability of tachyons: in the Minkowski space-time a tachyon is classically unstable under small perturbations. At the quantum level this means that small vacuum fluctuations of the field will inevitably induce the rolling down of tachyons into the states with unboundedly increasing amplitudes which may end up with a chaos or the \emph{tachyon`s condensation}.

In cosmology, the tachyon`s instability can be partially avoided by the specific choices of the energy momentum tensor
together with the introduction of various non-minimal couplings of the tachyon. An extensive literature is devoted to this subject, so that we shall mention here a few works \cite{Tahyon-cosmology}, referring readers to further literature mentioned in \cite{Tahyon-cosmology}.
At the same time, the G\"{o}del solution \cite{Godel:1949ga} can describe the rotating Universe, and although the original G\"{o}del metric does not reproduce the Hubble expansion, a number of its generalizations (including non-static ones) were found, which could be better approximation to the current observations.

In this letter, we shall show that minimally coupled tachyonic scalar or Dirac fields are classically stable in the rotating Universe if the absolute value of the field mass is small enough. By considering a family of  G\"{o}del-like solutions, describing both causal and non-causal rotating Universes, as well as various number of space-time dimensions (four and five), we are trying to show that the stability of tachyon is apparently not connected with such issues as space-time causality or a particular type of the rotating cosmological model, but seems to be appropriate to rotating universes in general.

Let us mention here a few previous papers which are closely connected to the present letter. The first consideration of the massless scalar field in the four dimensional non-causal G\"{o}del Universe \cite{Godel:1949ga} was by Hiscock \cite{Hiscock:1978iq}. Then, the above analysis was generalized by Leahy \cite{Leahy:1982dj} to scalar and spinor fields. In \cite{GuhaThakurta:1980ax} Thakurta considered massless scalar field in more general $D=4$ rotating cosmological backgrounds given by \cite{Ozsvath} and \cite{Vaidya:1967zz}. Finally, Radu \cite{Radu:1998yz} considered charged massive scalar field in the above noncausal G\"{o}del -like space-times.
Recently, there has been considered the quasinormal modes of massive scalar field for black holes immersed in
the 5-dimensional G\"{o}del space-time \cite{Konoplya:2005sy},\cite{Konoplya:2011ig},\cite{Konoplya:2011it}.
Although the calculations in all of the above cases are naturally very similar to ours, no tachyonic effects were considered in any of those works, which implied "normal" sign of the $\mu^2$-term. In the present work we  extend the above results through the consideration of more general cosmological backgrounds (both causal and non-causal four dimensional G\"{o}del-like space-times and five dimensional non-causal ones) and inclusion of the massive scalar and spinor fields. Yet, technically this generalization is quite simple and the main result here is the observation and interpretation of the classical (in)stabilities of massive fields in terms of tachyonic particles.

The paper is organized as follows: Sec II is devoted to perturbations of the test massive scalar field with the opposite (tachyonic) square mass term in the four dimensional G\"{o}del-like cosmological background. Sec III is devoted to discussion of the five-dimensional case which appears in the supergravity. In Sec. IV stability of the massive Dirac field in the G\"{o}del-like space-times is analyzed and the implication to the would-be superluminal neutrinos is discussed. In Sec V we discuss the obtained results.

\section{4-dimensional rotating Universe}

Let us first consider a two-parametric cosmological solution suggested in \cite{Vaidya:1967zz}, \cite{Ozsvath}, \cite{Tiomino} which is the generalization of the classical G\"{o}del metric \cite{Godel:1949ga}. The metric has the form
\begin{equation}
d s^2 = dt^2 - dx^2 - dz^2 + \frac{A^2 + \lambda/2}{A^2 + \lambda} e^{2 \alpha x} d y^2 + 2 e^{\alpha x} d y d t,
\end{equation}
where $\alpha = \sqrt {2 \lambda}$. The G\"{o}del metric \cite{Godel:1949ga} is reproduced in the limit $A=0$ and $\lambda = \Omega^2$.
The perturbation equation for the massive scalar field in a curved space-time can be written as
\begin{equation*}\label{scalar}
\square\Phi + \mu^2 \Phi= \frac{\partial}{\partial x^{\mu}}\left(g^{\mu\nu}\sqrt{-g}\frac{\partial \Phi}{\partial x^{\nu}}\right)  + \mu^2 \Phi =0,
\end{equation*}
where $\mu^2 > 0$ corresponds to the "normal" matter and $\mu^2 < 0$ to the tachyonic scalar field. The inverse metric and the determinant has the following form:
\begin{equation}
g^{00} = -1 - \frac{2 A^2}{\lambda}, \quad g^{11} = g^{33} = -1, \quad g^{02} = g^{20} = 2 e^{- \alpha x} \left(\frac{A^2}{\lambda} +1 \right),
\end{equation}
\begin{equation}
g^{22} = -2 e^{- 2 \alpha x} \left(\frac{A^2}{\lambda} +1 \right), \quad \sqrt{-g} = e^{\alpha x} \sqrt{\frac{\lambda}{2(A^2 + \lambda)}}.
\end{equation}
As the background metric possesses the Killing vectors $\partial_t$, $\partial_y$, $\partial_z$,
one can use the following anzats for the perturbation,
\begin{equation}\label{anzats}
\Phi(t, x, y, z) \sim e^{- i (\omega t - k_{2} y - k_{3} z)} \Phi(x).
\end{equation}
Then, the scalar field equation (\ref{scalar}) takes the form, which is similar to the one obtained in \cite{GuhaThakurta:1980ax},
\begin{equation}
\Phi,_{xx} + \alpha \Phi,_{x} - \left(k_{3}^2 +  \mu^2 + 2 k_{2}^2 e^{- 2 \alpha x} \left(\frac{A^2}{\lambda} +1 \right)
+ 4 k_{2} \omega e^{- \alpha x} \left(\frac{A^2}{\lambda} +1 \right) + \omega^2 \left(\frac{2 A^2}{\lambda} +1 \right) \right) \Phi =0.
\end{equation}
Going over to the new coordinate \cite{GuhaThakurta:1980ax},
\begin{equation}
u = \frac{2 |k_{2}| \sqrt{A^2 + \lambda}}{\lambda} e^{-\alpha x}, \quad \lambda \neq 0
\end{equation}
and making use of $\alpha = \sqrt {2 \lambda}$, we can re-write the wave equation as follows
\begin{equation}
\Phi,_{u u} - \left(\frac{1}{4} + \frac{k_{3}^2 + \mu^2 + (2 A^2 + \lambda) \omega^2/\lambda}{2 \lambda u^2} + \frac{k_2  \omega}{u \lambda |k_2|} \sqrt{A^2 + \lambda} \right) \Phi =0.
\end{equation}
Notice, that the $k_2 =0$ case does not give normalizable solutions which can be interpreted as perturbations. This can easily be shown in the same way as in \cite{Hiscock:1978iq}.

The general solution to the above equation can be written in terms of the confluent hypergeometric functions
\begin{equation}
\Phi = e^{-u/2} u^{\beta} \left(C_1 M\left(\beta + \frac{k_2  \omega}{\lambda |k_2|} \sqrt{A^2 + \lambda}, 2 \beta, u \right) + C_2 U\left(\beta + \frac{k_2  \omega}{\lambda |k_2|} \sqrt{A^2 + \lambda}, 2 \beta, u \right)\right),
\end{equation}
where $$\beta  = 1+ \sqrt{1 + 2 \frac{k_{3}^2 + \mu^2 + (2 A^2 + \lambda) \omega^2/\lambda}{\lambda u^2}}.$$
For the solution to be everywhere bound, i.e. to be treated as a perturbation, the following relation must be satisfied,
\begin{equation}
\frac{1}{2} + \left(\frac{1}{4} + \frac{k_{3}^2 + \mu^2 + (2 A^2 + \lambda) \omega^2/\lambda}{2 \lambda}\right)^{1/2} + \frac{k_2  \omega}{ \lambda |k_2|} \sqrt{A^2 + \lambda} = -n.
\end{equation}
Then, the wave function $\Phi$ is normalizable (that is the integral of energy over the whole space is finite) and the frequency is
\begin{equation}\label{omega}
\omega = - \frac{k_3}{|k_{3}|} \sqrt{A^2 +\lambda} \left((2 n+1) \pm \sqrt{(2 n+1)^2 - \frac{2 \lambda}{A^2 + \lambda} \left(n^2 + n
-\frac{k_{3}^2 + \mu^2}{2 \lambda}\right)} \right).
\end{equation}
According to (\ref{anzats}), complex values of $\omega$ with $Im \omega > 0$ means unboundedly growing mode, that is, the \emph{instability}.
The lowest mode $k_3 = n=0$ implies that $\omega$ is real in the region
\begin{equation}
1 + \frac{\mu^2}{A^2 + \lambda} > 0.
\end{equation}
The latter allows for the stable negative values of $\mu^2$, while
\begin{equation}
|\mu^2| < A^2 + \lambda.
\end{equation}
For the classical G\"{o}del solution we have $\lambda = \Omega^2$ and $A=0$, so that the condition for the tachyon`s stability reads
\begin{equation}
|\mu| < \Omega
\end{equation}
Alternatively, one can consider $\lambda$ to be the cosmological constant, then the matter density $4 \pi \rho = A^2 + \lambda$ \cite{Ozsvath}, \cite{Vaidya:1967zz}, so that
\begin{equation}
|\mu| < \sqrt{4 \pi \rho}.
\end{equation}

It is well known that the classical G\"{o}del metric ($A=0$, $\lambda = \Omega^2$) allows for the closed time-like curves, what makes this metric rather exotic. It would be, therefore, natural to connect the tachyon`s stability with this non-causality of asymptotic regions of the G\"{o}del space-time.
However, we shall show here that the stability bound exists also for the causal metrics with the same symmetry.

The metric (1) can be re-written in the cylindrical coordinates $(t', r, \phi, z)$ as follows
\begin{equation}
d s^2 = - d r^2 - d z^2 - \frac{\sinh^2(\lambda r)}{\lambda^2} d \phi^2 + \left(\frac{4 \Omega}{\lambda^2} \sinh^2(\lambda r/2) d \phi + d t' \right)^2.
\end{equation}
If
\begin{equation}
\frac{4 \Omega}{\lambda^2} \sinh^2(\lambda r/2) \left(1+ \left(1- \frac{4 \Omega^2}{\lambda^2}\right) \sinh^2(\lambda r/2)\right) >0,
\end{equation}
or briefly
\begin{equation}
\lambda > 2 \Omega,
\end{equation}
no closed time-like curves are allowed for the space-time. This is an example of the causal cosmological solutions suggested in \cite{Tiomino}.
However, the above derivation of the tachyon`s stability gap does not depend on the particular values of $\lambda$, so that for the generic metric of the form
\begin{equation}\label{metric18}
d s^2 = dt^2 - dx^2 - dz^2 + C e^{2 \alpha x} d y^2 + 2 D e^{\alpha x} d y d t,
\end{equation}
the tachyon is stable, if
\begin{equation}
|\mu^2| < \frac{D^2 \alpha^2}{4 |C- D^2|}.
\end{equation}
The latter formula includes both causal ($C = 0$) and non-causal ($C > 0$) space-times.
The causal $C<0$ case allows for the instability at sufficiently high overtones $n$, as can be seen for instance
from eq. (\ref{omega}) when $2 \lambda/(A^2 + \lambda) >4$, what happens if $A^2 <0$.
Yet, this instability happens also for massless particles, that is for the normal matter, what makes this particular case quite exotic.
Therefore, the stability gap for the tachyon exists for both non-causal and not exotic causal space-times.

\section{5-dimensional Universe with the gauge field}

In order to check if the existence of the stability gap for tachyons depends on the number of space-time dimensions,
we shall consider another G\"{o}del -like solution which has been recently actively studied in the context of supergravity and string theory.

The bosonic fields of the minimal (4+1)- supergravity theory consist
of the metric and the one-form gauge field, which are governed by the
following equations of motion
\begin{equation}\label{raz}
R_{\mu \nu} =2 \left(F_{\mu \alpha} F_{\nu}^{\alpha} -\frac{1} {6} g_{\mu \nu}
F^{2}\right)
\end{equation}
\begin{equation}\label{dva}
D_{\mu} F^{\mu \nu} =  \frac{1}{2 \sqrt{3}} \varepsilon^{\alpha \lambda
\gamma \mu \nu}  F_{\alpha \lambda} F_{\gamma \mu}
\end{equation}
Here, we have $ \varepsilon_{\alpha \lambda \gamma \mu \nu} = \sqrt{-g} \epsilon_{\alpha \lambda \gamma \mu \nu}$.

In the Euler coordinates $(t, r, \theta, \psi, \phi)$, the solution of the
equations of motion (\ref{raz}), (\ref{dva}), describing the G\"{o}del universe, has the form \cite{Gimon:2003ms}:
\begin{eqnarray}
ds^2 &=& - (dt + jr^2 \sigma_{L}^{3})^2 + dr^{2} \nonumber\\
&&+\frac{r^2}{4}(d \theta^{2} + d \psi^{2} + d \phi^{2} + 2 \cos \theta d \psi d \phi),
\end{eqnarray}
where $\sigma_{L}^{3}= d \phi + cos \theta d \psi$. The parameter $j$, similarly to the parameter of
the universe angular momentum $\Omega$, defines the scale of the G\"{o}del background.  At $j=0$ we have
the Minkowski space-time.

As it was shown in \cite{Konoplya:2011ig}, the scalar field equation can be written in the following form
\begin{equation}\label{nobheq}
\frac{d^2 R}{dr^2} + \left((1 - 4 j^2 r^2)\omega^2 - 8 j m \omega- \frac{\frac{3}{4} + 4 \lambda}{r^2} - \mu^2)\right) R =0.
\end{equation}

Making use of $\lambda=\ell(\ell+1)$ and introducing the new function $p(r)$,
$$R=\left\{\begin{array}{ll}e^{-j\omega r^2}r^{-2\ell-1/2}p(r), & \re{j\omega}>0; \\e^{j\omega r^2}r^{-2\ell-1/2}p(r), & \re{j\omega}<0.\end{array}\right.
$$
we can further reduce the wave equation to the Kummer's form
\begin{equation}
zp''(z)+(b-z)p'(z)=ap(z),
\end{equation}
with respect to the new coordinate $$z=\left\{\begin{array}{ll}2j \omega r^2, & \re{j\omega}>0; \\-2j \omega r^2, & \re{j\omega}<0.\end{array}\right.$$
Here $b=-2\ell$ and
$$8j\omega a=\left\{\begin{array}{ll}-8j(\ell-m)\omega-\omega^2+\mu^2, & \re{j\omega}>0; \\-8j(\ell+m)\omega+\omega^2-\mu^2, & \re{j\omega}<0.\end{array}\right.$$
The solution to the Kummer's equation, which corresponds to the regular at the origin ($r=0$) wave function $\Psi(r)$, is
$$p(z)\propto z^{1+2\ell} M(a+2\ell+1,2\ell+2,z).$$
Here $M(A,B,z)$ is the generalized hypergeometric series, which has an irregular singularity at $z=\infty$,
unless $A$ is a non-positive integer. Thus, one has
$$a=-n-2\ell, \qquad n=1,2,3\ldots .$$
This gives us the expression for $\omega$:
\begin{equation}\label{godelqnms}
\omega=\pm4j\left(n+\ell\pm m+\sqrt{(n+\ell\pm m)^2+\frac{\mu^2}{16j^2}}\right).
\end{equation}
The lowest mode $n=1$, $\ell = m = 0$ gives us the condition for the tachyon stability
\begin{equation}
|\mu| < 4 j.
\end{equation}

\section{Remark on superluminal neutrino}

In addition to a massive scalar field, let us consider the Dirac field in the background (\ref{metric18}).
For massive Dirac fields in a curved background
$g_{\mu \nu}$, the equation of motion reads \cite{Ivanenko}
\begin{equation}\label{em:Dirac}
(\gamma^ae_a^{~\mu}(\pd_\mu+\Gamma_\mu)+\mu)\Phi=0, \quad ( s =
\pm 1/2)
\end{equation}
where $\mu$ is the mass of the Dirac field, and $e_a^{~\mu}$ is the
tetrad field, defined by the metric $g_{\mu\nu}$:
$$ g_{\mu\nu}=\eta_{ab}e_\mu^{~a}e_\nu^{~b}, \quad
g^{\mu\nu}=\eta^{ab}e_a^{~\mu}e_b^{~\nu}, $$
\begin{equation}
\quad e_a^{~\mu}e_\mu^{~b}=\delta_a^b, \quad e_\mu^{~a}e_a^{~\nu}=\delta_\mu^\nu,
\end{equation}
where $\eta_{ab}$ is the Minkowski metric, $\gamma^a$ are the Dirac
matrices: $$\{\gamma^a,\gamma^{b}\}=2\eta^{ab},$$ and $\Gamma_\mu$ is the spin
connection  \cite{Ivanenko},
\begin{equation}
\Gamma_\mu=\frac{1}{8}[\gamma^a,\gamma^b]~g_{\nu\lambda}~e_a^{~\nu}~e_{b~;\mu}^{~\lambda}.
\end{equation}
Using the following tetrads:
\begin{equation}
e_{0}^{(0)}=e_{1}^{(1)} = e_{3}^{(3)} =1, \quad e_{2}^{(0)} = D e^{\alpha x}, \quad e_{2}^{(2)} = \sqrt{C- D^2} e^{\alpha x},
\end{equation}
where $(i)$ are tetrad indices, and taking $D=1$ in (\ref{metric18}), one can calculate the spinor connections:
$$ \Gamma_0=\frac{\alpha}{8 \sqrt{1-C}}[\gamma^1,\gamma^2], \quad
\Gamma_1=\frac{e^{\alpha x}}{8 \sqrt{1-C}} [\gamma^0,\gamma^2],$$
\begin{eqnarray}
\Gamma_2=-\frac{\alpha e^{\alpha x}}{8}\left([\gamma^0,\gamma^1]+\frac{1 - 2 C}{\sqrt{1-C}}[\gamma^1,\gamma^2]\right), \quad
\Gamma_3=0.
\end{eqnarray}
As for the scalar field, we shall consider the adiabatic approximation for the Universe's expansion and assume the following ansatz
\begin{equation}\label{anzats}
\Phi(\tau,\eta,y,z)=e^{-\imo\omega\tau+\imo k_2 y+\imo k_3 z}\Psi(\eta),
\end{equation}
where $\eta = e^{- \alpha x}$ is a new spatial variable.  
Replacing $\Phi$ in the Dirac wave equation by (\ref{anzats}) and choosing 4-component spinor as
\begin{equation}
\Psi(\eta)=\left(1+\frac{{\tilde k}_3}{\mu}\gamma^3\right)\left(
                                                               \begin{array}{c}
                                                                 \psi_1(\eta) \\
                                                                 \psi_2(\eta) \\
                                                                 \psi_3(\eta) \\
                                                                 \psi_4(\eta) \\
                                                               \end{array}
                                                             \right),
\qquad {\tilde k}_3 = k_3\left(\sqrt{1+\frac{\mu^2}{k_3^2}}-1\right),
\end{equation}
we can reduce the matrix equation to the second-order differential equations for each component
\begin{equation}\label{Dirac-wavelike}
\psi''(\eta)-\left(\frac{k_2^2}{1-C}+\frac{2k_2{\omega}\pm k_2\sqrt{1-C} \alpha}{(1-C)\eta}+\frac{4F(k_3)^2-\alpha^2}{4\eta^2}+\frac{C{\omega}^2}{(1-C)\eta^2}\right)\frac{\psi(\eta)}{\alpha^2}=0,
\end{equation}
where $ m = sgn(k_2)$, and
$$F(k_3)=k_3\frac{\mu^2-{\tilde k}_3^2}{\mu^2+{\tilde k}_3^2}+2{\tilde k}_3\frac{\mu^2}{\mu^2+{\tilde k}_3^2}+\frac{\alpha}{4 \sqrt{1-C}}.$$
In the limit $\mu\rightarrow0$ we reproduce the formula (102) of \cite{Leahy:1982dj} (one should take $C=1-\alpha=1/2$).

Here one should care about $k_2 =0$ mode which is, although being the lowest frequency mode, is not normalizable, so that it cannot influence the stability analysis. The above wave-like equations can be reduced to the same hypergeometric form as the scalar field, though with different coefficients. Therefore, one can obtain the explicit expression for the frequency,
\begin{equation}\label{sqaure-root2}
\omega=-\frac{(2n+1\pm1)\alpha}{2\sqrt{1-C} m}\pm\frac{\sqrt{(2n+1\pm1)^2 C\alpha^2 /(1-C)+4F^2(k_3)}}{2}.
\end{equation}
The ``-'' sign for $n=0$ violates conditions of convergence of the Dirac field norm (see \cite{Leahy:1982dj} for details), so that we are able to write both solutions as
\begin{equation}
\omega=-\frac{(n+1)\alpha}{\sqrt{1-C} m}\pm\sqrt{(n+1)^2 C \alpha^2/(1-C)+F^2(k_3)}.
\end{equation}
In the limit $k_3 =0$, one has $F(k_3) = \mu + \alpha/4 \sqrt{1-C}$, so that the expression under the square root has always non-vanishing imaginary part, which means the instability under the "-" sign in front of the square root. Thus, Dirac tachyons are unstable in the G\"{o}del Universe for arbitrary masses. 
However, the stability can be achieved if one considers not only rotating, but also expanding Universe, which agrees with the current observations. 
Thus, for a G\"{o}del-like metric with expansion, given by the metric
\begin{equation}
ds^2=\left(dt+a(t)e^{\alpha x}dy\right)^2-a(t)^2\left(dx^2+B^2e^{2\alpha x}dy^2+dz^2\right),
\end{equation}
we find (\cite{RK-forth}) that the tachyon is stable if
\begin{eqnarray}
|\mu|&\leq \frac{3H}{2}\sqrt{1+\frac{16(1-B^2)}{1+18B^2/\Omega_H^2}}, \quad B \neq 1.
\end{eqnarray}
Taking account of the modern upper bound on the Universe rotation $\Omega_H \simeq 74$, one can find the bound on mass as,
$$|\mu|\lesssim\frac{9H}{2}.$$
The above bound is three times larger than the one which comes from the pure Freedman Universe, $|\mu|\lesssim\frac{3 H}{2}$, which means that the rotation can be a dominating (over expansion) factor for the upper bound for the tachyon`s mass.

Therefore, we could suppose that \emph{in G\"{o}del-like rotating and expanding cosmological backgrounds (either non-exotic causal or non-causal)
tachyonic particles of various spin are stable, if}
\begin{equation}
|\mu| \lesssim C(\Omega),
\end{equation}
where $\Omega$ is positive constant describing the scale of the rotating Universe, which can usually be associated with the angular momentum.
The constant $C > 0$ depends on the particular choice of the rotating cosmological model and on the spin of the field.

Now let us make some "naive" estimates for the upper bound of the tachyon mass, starting from the upper bound on the rotation rate of the Universe, which is $\Omega \lesssim 74 H$ according to \cite{Obukhov}, where $H$ is the Hubble constant. Going over from the geometrical units $c=\hbar = G=1$ to SI
in the inequality $\mu c/\hbar < \Omega$, we obtain the following upper bound for the tachyon`s mass
\begin{equation}
|\mu| \lessapprox 10^{-36} m_{e},
\end{equation}
where $m_{e}$ is the electron mass. Here we have considered rather not realistic cosmological model without expansion.
Yet, it will be shown in our forthcoming paper \cite{RK-forth} that the wide class of the G\"{o}del-like Universes with
expansion produces the upper bound on the tachyon`s mass exactly of the same order. This happens because the wave equation for the scalar and spinor fields can be again reduced to a similar hypergeometric form. Moreover, the non-rotating expanding Universe itself allows for the stability of tachyons, though with a few orders smaller upper bound for the mass \cite{RK-forth}.

A crucial and still an open question is what the experimentally witnessed lower bound for the neutrino`s mass is?
To the best of our knowledge, there is no generally accepted lower bound on the muon neutrino mass. Instead,
the difference of squared mass between various eigenstates of the neutrino oscillations were detected. For instance,
it is supposed that there is at least one neutrino mass eigenstate with a mass of at least $0.04 eV$ \cite{PartRev}.
The found here stability bound is very small, though it could be further improved by consideration
cosmological models which are better approximations to the current observation than
the Friedman-G\"{o}del-like universes considered here.
If the lower bound on the neutrino`s mass in the OPERA experiment is as large as it is expected, then,
our result should signify against the tachyonic nature of the neutrino.

\section{Conclusions}

We tried to demonstrate here that acceptance of a simple concept of the Universe rotation leads to the stability of tachyons.
In this letter we have shown that the tachyon, described by either massive scalar or Dirac fields, is stable in the wide class of rotating G\"{o}del-like cosmological backgrounds, if mass of the tachyon is small enough.
By considering causal and non-causal G\"{o}del-like metrics and various number of space-time dimensions, we intended to show that the existence of the stability region for tachyons is not determined by such exotic issues as the (non)-causality of the space-time or existence of extra spatial dimensions. Although the current damped rotation rate of the Universe cannot be large, the larger rotation rate of the early Universe could allow for much larger masses of tachyons in the past.

We have shown that in the rotating Universe, tachyons with very small masses can be stable, and this is remarkable that the upper bound on the tachyon`s mass (although it accurate value depends on the concrete choice of the cosmological model) is many orders less than the electron`s mass. Potentially, this might be an explanation why only the lightest particles, such as neutrinos in the recent OPERA experiment \cite{OPERA:2011zb}, have a chance to be tachyons.
Yet, the lower bound of the neutrino`s mass in the OPERA is apparently much larger than the stability bound, which means that our result signifies against the tachyonic nature of the neutrinos.

Let us note that the situation with the small gap of stability of the tachyon is very similar to the well-known
Breitenlohner-Freedman bound which occurs in the AdS space-time \cite{Breitenlohner:1982bm}. Indeed, the similarities between the spectrum of normal modes of the G\"{o}del space-time and the AdS one have been recently noticed in \cite{Konoplya:2011ig}, \cite{Konoplya:2011it}. Thus, the quasinormal modes of a black hole in AdS space-time approach the normal modes of pure AdS space-time in the limit of vanishing black hole radius \cite{Konoplya:2002zu}, \cite{Konoplya:2011qq},  and the same is true for the G\"{o}del space-time, where the universe scale $j$ plays a similar role to the anti-de Sitter radius $R$ in AdS. Therefore, the existence of the analog of the Breitenlohner-Freedman bound could be expected.
An existence of the observed similarities in the spectra of the asymptotically G\"{o}del and anti-de Sitter space-times might be due to the fact that G\"{o}del metric belongs to the common family of space-times which can be interpreted as squashed anti-de Sitter geometries \cite{Rooman:1998xf}.

An important issue that was beyond our consideration is the quantization of tachyons, which is important for making practical estimations for such quantity as the vacuum energy of the field. The above discussed normal modes of the tachyonic fields, although are normalizable, do not form a complete set, what was noticed first time in \cite{Leahy:1982dj} for massless fields. Therefore, quantization for such systems deserves a separate consideration.

\section*{Acknowledgments}
This work was supported by the European Commission through the Marie Curie International Incoming Contract.
I thank B. I. Konoplya, who brought my attention to this problem. Also i would like to acknowledge useful discussions with A. Zhidenko.

\end{document}